\documentclass[namedreferences]{kluwer}

\usepackage{klups}

\usepackage[dvips]{graphicx}

\usepackage{makeidx}
\def\indexname{Index}
\makeindex

\begin{document}
\begin{article}

\begin{opening}
\title{Multiwavelength Study Of Two Unidentified $\gamma$-Ray Sources}
\vspace{-0.5 cm}
\author{Nicola \surname{La Palombara}, Patrizia \surname{Caraveo}}
\institute{IASF/CNR - Sezione di Milano `G.Occhialini', Via E.Bassini 15, I--20133 Milano (I)}
\author{Roberto \surname{Mignani}}
\institute{European Southern Observatory, Karl Schwarzschild Strasse 2, D--85740 Garching (D)}
\author{Evanthia \surname{Hatziminaoglou}}
\institute{Instituto de Astrofisica de Canarias, Via Lactea, E--38200 La Laguna--Tenerife (E)}
\author{Giovanni F. \surname{Bignami}}
\institute{CESR, 9 Avenue du colonel Roche, F--31028 Toulouse (F)}
\author{Mischa \surname{Schirmer}}
\institute{Isaac Newton Group of Telescopes, Edificio Mayantigo, Calle Alvarez Abreu 68, E--38700 Santa Cruz de la Palma (E)}
\runningauthor{Nicola La Palombara et al.}
\runningtitle{Multiwavelength Study Of Two Unidentified $\gamma$-Ray Sources}

\vspace{-0.5 cm}
\begin{abstract}
Most counterparts of the identified low--latitude $\gamma$--ray sources
are {\it Isolated Neutron Stars} (INSs).  Since INSs are characterized by an extremely
high value of $f_{X}/f_{opt}$, a systematic X--ray/optical coverage of the
fields of unidentified low--latitude $\gamma$--ray sources is the best way to
unveil INS counterparts of unidentified sources.  Since the low--latitude
sources are heavily affected by the interstellar absorption in both the X-ray
and optical bands, we decided to apply the above strategy to two
middle--latitude {\it EGRET} sources, which could belong to a local galactic
population:  3EG J0616-3310 and 3EG J1249-8330.  Here we report on the global
X-ray characterisation of about 300 objects, on their candidate
optical counterparts and on the preliminary results of their identification.
\end{abstract}

\end{opening}
\vspace{-1 cm}


\section {X--ray observations, data reduction and source detection}
\vspace{-0.3 cm}

\begin{figure}
\centering
\resizebox{\hsize}{!}{\includegraphics[angle=+90,clip=true]{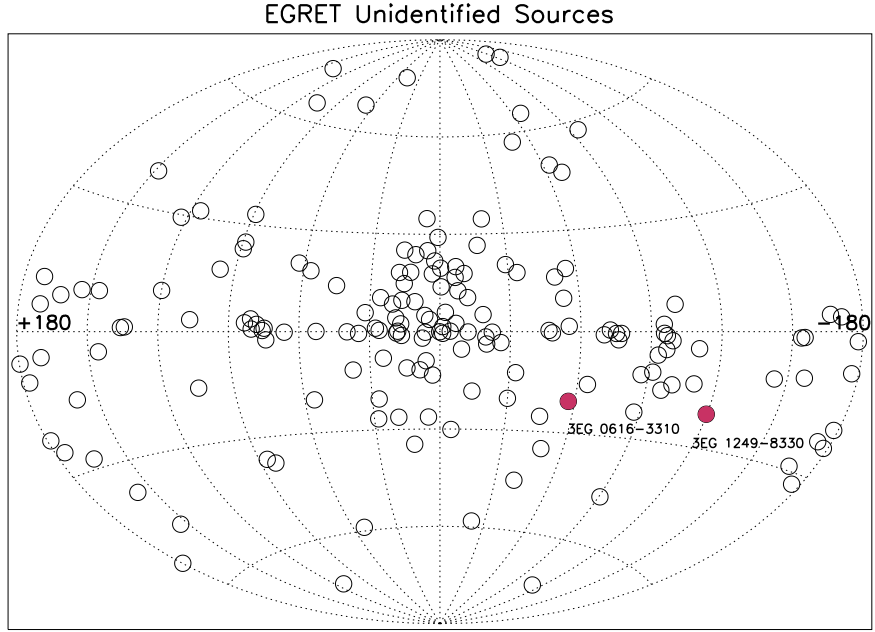}}
\caption{Sky distribution of the unidentified {\em EGRET} sources:  the two red circles show the
celestial position of 3EG J0616-3310 and 3EG J1249-8330 (courtesy of S.
Vercellone)}\label{galacticposition}
\end{figure}

3EG J0616-3310 and 3EG J1249-8330 are two middle--latitude ($b \simeq -20^\circ$), still
unidentified, {\em EGRET} sources (Fig.~\ref{galacticposition}).  Since they have no candidate
extra--galactic counterparts, we assumed that {\em Isolated Neutron Stars} (INSs) are the most
likely sources of the $\gamma$--ray emission.  To this aim, we performed a multi--wavelength search
of celestial objects with high (i.e.  $\ge$1000) X--ray--to--optical flux ratios, according to an
investigation strategy which was already successfully used for the identification of {\em Geminga}
\cite{Bignami} and of other {\em EGRET} sources \cite{Mirabal,Caraveo2001}.

\begin{figure}[h]
\tabcapfont
\centerline{%
\begin{tabular}{c@{\hspace{1pc}}c}
\includegraphics[width=2.5in]{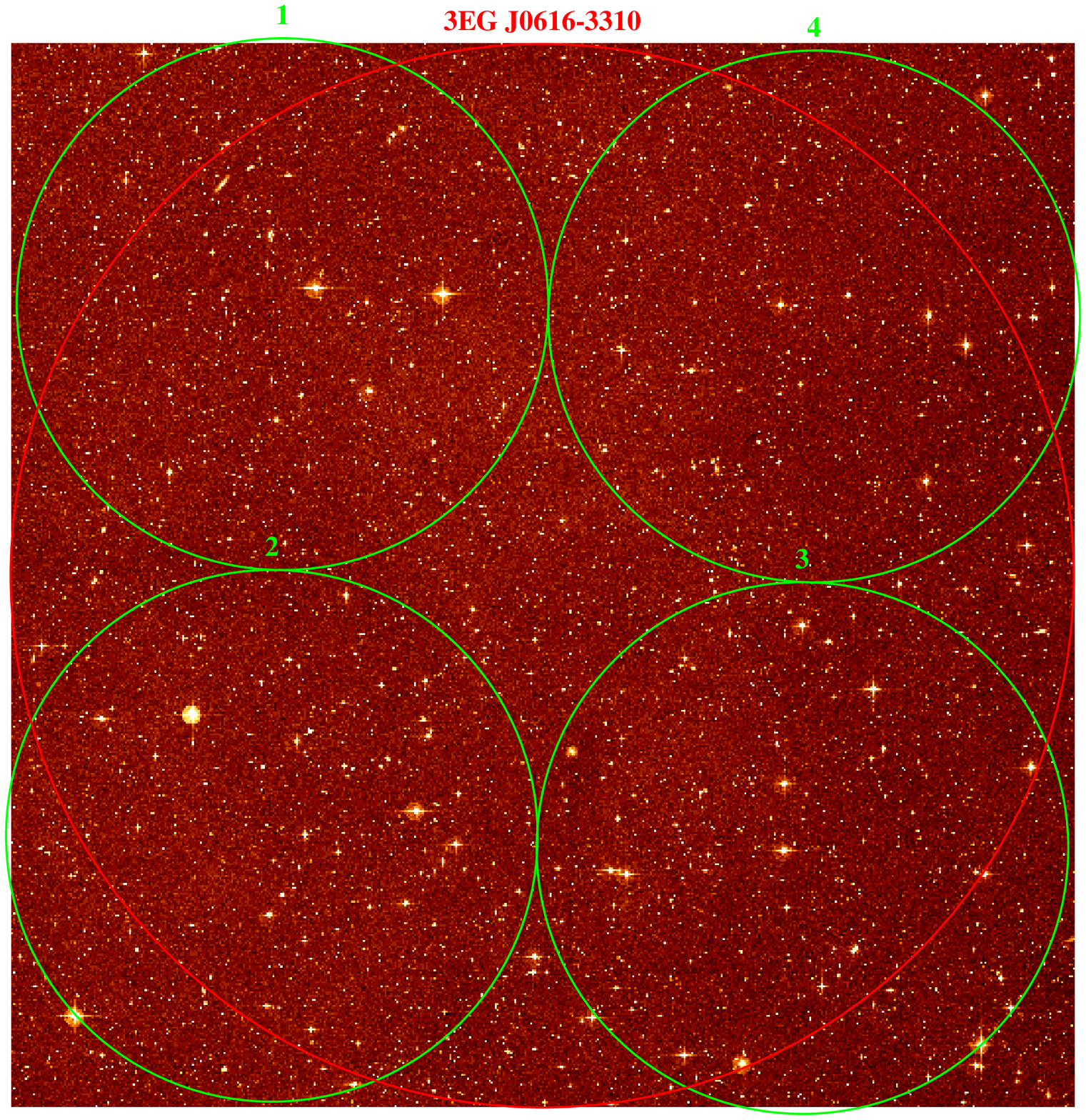} &
\includegraphics[width=2.5in]{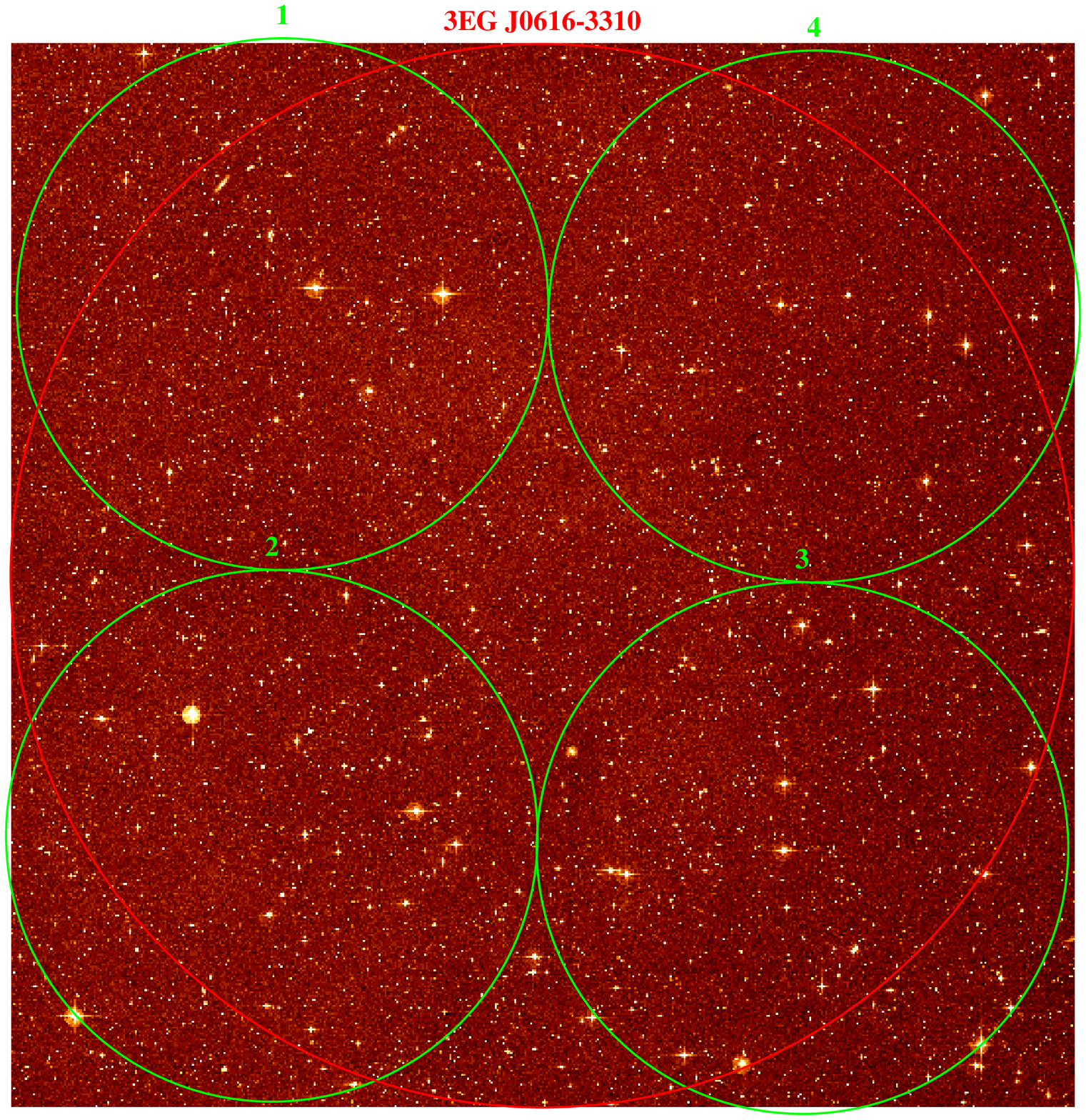} \\
\end{tabular}}
\caption{Optical image of the sky--area of 3EG J0616-3310 (left) and 3EG J1249-8330
(right), with the overlay of the $\gamma$--ray  error--box  (large circle) and the
field--of--view   covered  by  the  four  {\em  XMM--EPIC}   pointings   (small
circles)}\label{errorbox}
\vspace{-0.5 cm}
\end{figure}

Both sources are positioned within error circles of $\sim0.5^\circ$ radius which
corresponds roughly to the diameter of the {\em EPIC} field--of--view, the focal plane
camera of the {\em XMM--Newton} observatory.  Each region was covered by four 10 ks {\em
EPIC} pointings (Fig.~\ref{errorbox}) and each of the eight pointings includes three
data-sets for the {\em PN}, {\em MOS1} and {\em MOS2} cameras \cite{Struder,Turner},
respectively.  All data-sets were independently processed with the standard {\em
XMM-Newton Science Analysis System} (SAS).

After the standard processing pipeline, we performed the source detection in 7
different energy ranges. Namely, we considered two coarse soft/hard bands
(0.5--2 and 2--10 keV) and a finer energy division (0.3--0.5, 0.5--1, 1--2,
2--4.5, 4.5--10 keV).

To increase the {\em signal--to--noise} (S/N) ratio and to reach fainter X--ray flux limits, we
analysed together the events from the three instruments, then we selected only the sources with a
{\it detection likelihood} $L>8.5$ in at least one of the predefined energy ranges.  With this
criterion we found a total of 146 sources in the 3EG J0616-3310 error circle and 148 sources in the
3EG J1249-8330 one \cite{La Palombara}.
\vspace{-0.5 cm}


\section{X--ray results}
\vspace{-0.3 cm}
\subsection{Full statistics}
\vspace{-0.3 cm}

In Table~\ref{detections} we report the number, as well as the percentage, of
the sources detected in each energy range.
\vspace{-1 cm}

\begin{center}
\begin{table}[h]
\caption{X--ray sources detected in each energy range.}\label{detections}
\footnotesize{
\begin{tabular}{|c|c|c|} \hline
Source 		& 3EG J0616-3310	& 3EG J1249-8330	\\ \hline
Range (keV)	&	N(\%)	&	N(\%)	\\ \hline
0.5--2		& 119 (81.5)	& 119 (80.4)	\\
2--10		& 41 (28.1)	& 42 (28.4)	\\ \hline
0.3--0.5		& 28 (19.2)	& 14 (9.5)		\\
0.5--1		& 73 (50)		& 43 (29.1)	\\
1--2		& 81 (55.5)	& 77 (52)		\\
2--4.5		& 47 (32.2)	& 36 (24.3)	\\
4.5--10		& 4 (2.7)		& 8 (5.4)		\\ \hline
Total		& 146		& 148		\\ \hline
\end{tabular}}
\vspace{-1 cm}
\end{table}
\end{center}

The number of detected sources within each field is strongly dependent on the selected
energy band.  More than 80\% of all sources are detected in the range between 0.5 and 2
keV, with more than 50 \% of the total detected in the sub--range 1--2 keV.  However, only
very few sources are detected at very high or very low energies.  It is also interesting
to remark that the percentage of sources detected below 1 keV is lower for the 3EG
J1249-8330 field than for the 3EG J0616-3310, whereas the number of sources is comparable
at higher energies.  Such a difference is very likely due to the column density of the
interstellar gas, which is higher in the first field ($\sim 2.5\cdot10^{20} cm^{-2}$ and
$\sim 10^{21} cm^{-2}$ for 3EG J0616-3310 and 3EG J1249-8330, respectively).

\subsection{Source distribution over the Count Rates}

\begin{figure}[h]
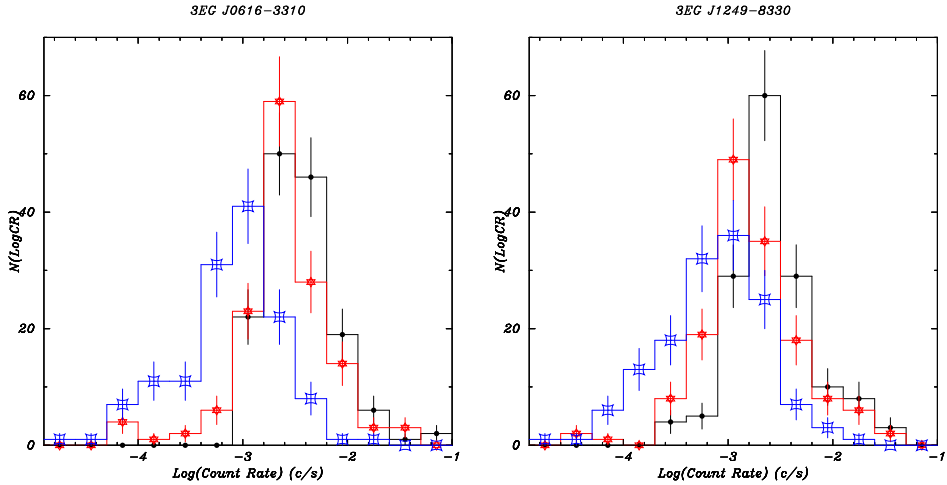

\tabcapfont
\centerline{%
\begin{tabular}{c@{\hspace{1pc}}c}
\includegraphics[width=2.5in,angle=-90]{cr_0616_3-new.ps} &
\includegraphics[width=2.5in,angle=-90]{cr_1249_3-new.ps} \\
\end{tabular}}
\caption{Histogram of the source distribution over the logarithm of their CR for the two whole
error--circles of 3EG J0616-3310 (left) and 3EG J1249-8330 (right), in the energy ranges 0.3--10 keV
(black line, filled circles), 0.3--2 keV (red line, small stars) and 2--10 keV (blue line, large squares)}\label{countrate}
\end{figure}

Fig.~\ref{countrate} shows the histogram of the source distribution over the logarithm of their {\em
Count Rate} (CR) in three energy ranges for the whole error--circle of the two {\em EGRET} fields.
Note that in the full energy range (i.e.  0.3-10 keV) the distribution peak is around logCR=-2.65
(i.e.  $CR\simeq2.2\cdot10^{-3}$ cts/s) for both error--circles.  Since the number of detected
sources for CR bin decreases below the peak CR, we assume that our sample becomes incomplete for
lower CR values:  therefore we consider the CR peak values as our completeness limit.

Moreover, we can see that, in the case of 3EG J0616-3310, the source distributions in the low and in
the full enegy range are very similar, since the peak count rate below 2 keV is equal to or just
below the full energy one; in the high energy range, however, the source distribution is very
different and most of the detected sources have count rates below $10^{-3}$ cts/s.  For 3EG
J1249-8330 there is a larger difference between the full and the low energy ranges, perhaps due to
the higher column density of the interstellar gas.
\vspace{-0.3 cm}

\subsection{Source distribution over the Signal--to--Noise ratio}
\vspace{-0.3 cm}

\begin{figure}[h]
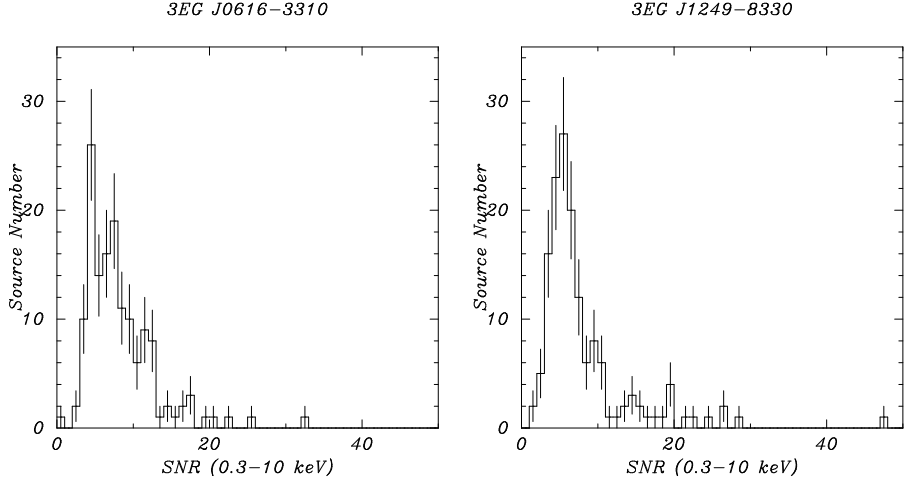

\tabcapfont
\centerline{%
\begin{tabular}{c@{\hspace{1pc}}c}
\includegraphics[width=2.5in,angle=-90]{figure2-0616NEW.ps} &
\includegraphics[width=2.5in,angle=-90]{figure2-1249NEW.ps} \\
\end{tabular}}
\caption{Histogram of the source distribution over their {\em S/N ratio} for 3EG J0616-3310 (left) and 3EG J1249-8330 (right)}\label{snr}
\end{figure}

In Figure~\ref{snr} we report the histogram of the source distribution as a function of
the S/N ratio (in the whole energy range) for the two {\em EGRET} fields.  The two
distributions peak at 4.5 and 5.5 for, respectively, the 3EG J0616-3310 and the 3EG
J1249-8330 sources.  Moreover, in the case of the 3EG J0616--3310 sample, a significant
fraction of the sources has a S/N ratio higher than 10; on the other hand, only one source
of the 3EG J1249-8330 field has a very high (47.16) S/N ratio.

\subsection{Source Hardness Ratios}
\vspace{-0.3 cm}
The count statistics of the sources in our sample is very low (usually of the order of a
few tens of photons) and therefore it is not possible to  accumulate  a significant
spectrum and to perform a spectral fit.  As an alternative,  we used the CR
in the seven energy ranges to calculate four different  {\em Hardness  Ratios} (HRs):
\begin{itemize}\vspace{-0.3 cm}
\item HR1=[CR(0.5-1)-CR(0.3-0.5)]/[CR(0.5-1)+CR(0.3-0.5)]\vspace{-0.3 cm}
\item HR2=[CR(1-2)-CR(0.5-1)]/[CR(1-2)+CR(0.5-1)]\vspace{-0.3 cm}
\item HR3=[CR(2-4.5)-CR(0.5-2)]/[CR(2-4.5)+CR(0.5-2)]\vspace{-0.3 cm}
\item HR4=[CR(4.5-10)-CR(2-4.5)]/[CR(4.5-10)+CR(2-4.5)]\vspace{-0.3 cm}
\end{itemize} 

\begin{figure}[h]
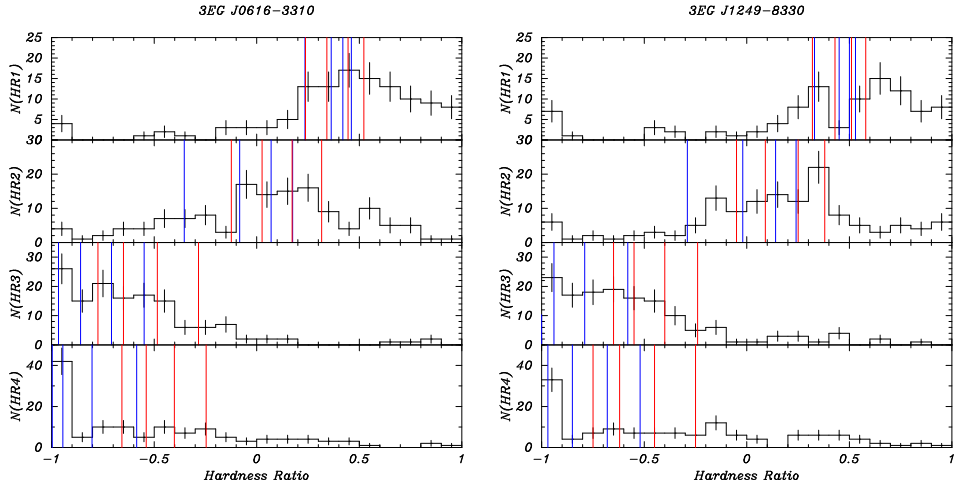

\tabcapfont
\centerline{%
\begin{tabular}{c@{\hspace{1pc}}c}
\includegraphics[width=2.5in,angle=-90]{0616_4hr_histo_simul_HK.ps} &
\includegraphics[width=2.5in,angle=-90]{1249_4hr_histo_simul_HK.ps} \\
\end{tabular}}
\caption{Histogram of the source distribution over their HRs for 3EG J0616-3310
(left) and  3EG J1249-8330  (right).  Blue bars indicate the expected HRs for
thermal {\em bremsstrahlung} spectra  with  kT=0.5, 1, 2 and 5 keV (from  left to  right).  Red bars
indicate the expected HRs for power--law spectra with $\alpha$=1, 1.5, 2 and 2.5
(from right to left).}\label{hr}
\end{figure}

Fig.~\ref{hr} illustrates the histograms of the number of detected sources as a function
of the single HRs.  From a global point of view, we found that most of the sources have
HR1>0, -0.2<HR2<+0.4, HR3<0 and HR4$\simeq$-1.  These results suggest that our source
population is characterised, on average, by rather soft spectra, with the count
distributions peaking between 0.5 and 2 keV.  In the majority of cases, the CR is higher
in the 1--2 keV range than in the 0.5--1 keV one, while several sources show a significant
CR up to 4.5 keV but only very few of them have a very soft or a very hard spectrum.

We also performed simulations to estimate the expected HRs for different spectral models.
For comparison, the results are shown as vertical bars in the Fig.~\ref{hr}.  It is
possible to see that the measured HR values are compatible with a rather wide range of
both temperatures and photon indexes, thus suggesting that we are sampling a variety of
types of serendipitous sources.  Such a result is not surprising, since the two areas
observed are at medium galactic latitude and therefore should contain both galactic and
extragalactic X--ray sources.  Making use of the results of the simulations, we could
tentatively assign to each source its most likely spectral parameters, both in the form of
temperature and photon--index.

\subsection{The Log(N)--Log(S) source distribution}
\vspace{-0.3 cm}
In order to evaluate the source fluxes, we considered the total count rate of
each source in the 0.3--10 keV energy range and assumed a template {\em
power--law} spectrum with photon--index $\alpha=1.7$.  As far as the hydrogen column
density $n_{\rm H}$, we considered the estimated mean total galactic value for
each of the eight EPIC pointings.  To compute the corresponding
count-rate-to-flux Convertion Factor ({\em CF}) for the eight individual
pointings, we considered separately the three {\em EPIC} focal plane cameras
and carefully combined both their {\em CFs} and their effective exposure times.

\begin{figure}[h]
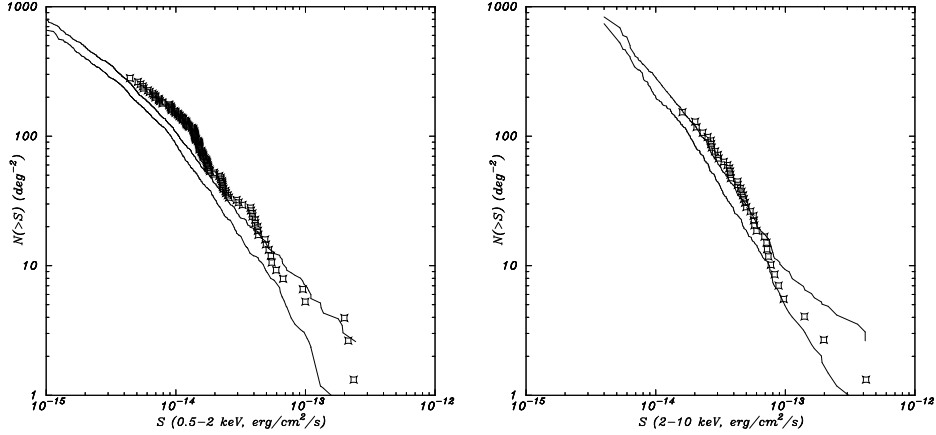

\tabcapfont
\centerline{%
\begin{tabular}{cc}
\includegraphics[width=2.25in,angle=-90]{lognlogs_05_2_0616_sqplot_all.ps} &
\includegraphics[width=2.25in,angle=-90]{lognlogs_2_10_0616_sqplot_all.ps} \\
\end{tabular}}
\caption{Cumulative Log(N)--Log(S) distribution of the detected sources in the
soft (0.5--2 keV, left) and hard (2--10 keV, right) energy ranges.  In both the diagrams the
two solid lines are the lower and upper limits of the high--latitude
Log(N)--Log(S)}\label{logNlogS}
\end{figure}

Fig.~\ref{logNlogS} shows the cumulative Log(N)--Log(S) distributions of the
sources detected in the two energy ranges 0.5--2 and 2--10 keV.  For comparison,
it shows also the lower and upper limits of the same distributions as measured
at high galactic latitude, where only extragalactic sources should be observed
\cite{Baldi}.  It is interesting to note that in the soft energy band the
source density is higher than at high latitudes. This indicates a significant
excess of galactic sources.  The same is not true in the hard energy range, where the
source distribution is in good agreement with the high--latitude one,
implying that galactic sources provide a negligible contribution at high energies.
\vspace{-0.5 cm}
\section{The optical analysis}
\vspace{-0.3 cm}
\subsection{Search for optical counterparts}\label{search}
\vspace{-0.3 cm}

In order to search for possible optical counterparts of our X--ray sources, we
cross--correlated their coordinates with the position of the sources of two
optical/infrared catalogues:  GSC2.2 (http://www--gsss.stsci.edu) and 2MASS
(http://pegasus.phast.umass.edu).  We found that only for very few cases an X--ray source
could be clearly associated with a single optical object:  we used the celestial positions
of these optical sources to perform the astrometric correction of all the X--ray source
coordinates, which turned out to be always lower than 2$''$.  Therefore we assigned a
conservative value of 5$''$ to our X--ray error circle radii.

Firstly, we considered the X--ray sources of all the eight pointings and selected, as
possible optical counterparts, all the GSC2.2 and 2MASS sources within their error
circles.  For some X--ray sources we found more than one candidate, while for several
sources we found no counterpart within 5$''$.  Such a result does not come as a surprise.
The GSC2.2 catalogue has a Bj limiting magnitude $\simeq$22.5 in the IIIaJ passband.
Taking into account both the completeness limit of the X--ray source detection and the
typical $f_{X}/f_{opt}$ ranges of the various classes of X--ray sources (section
~\ref{identification}), we estimated that the faintest counterparts should have V $\simeq$
22 for stars and V $\ge$ 25 for other classes of celestial objects.  We therefore expect
not to be able to detect the optical counterparts for several X--ray sources, for which we
assume a limiting magnitude Bj=22.5.

To reach deeper limiting magnitudes we used the {\em Wide Field Imager} of the MPG/ESO 2.2
m telescope at La Silla.  Three of the four {\em EPIC} pointings of 3EG J0616-3310 (i.e.
the pointings 2, 3 and 4 of Fig.\ref{errorbox}) were observed down to V$\simeq$25.5
\cite{Schirmer}.  Since our optical data should provide reliable counterparts for a large
fraction of our X--ray sources, we analysed them separately.  Even in this case, however,
we found no counterpart within 5$''$ for a few X--ray sources, for which we assumed a
limiting magnitude V=25.5.

\vspace{-0.3 cm}

\subsection{Source characterization and identification}\label{identification}
\vspace{-0.3 cm}

Based on the available X--ray and optical data, we performed a preliminary
characterization of all the X--ray sources in our sample.

From the X--ray point--of--view, we considered both the detection bands yielding
significant detection and, based on them, the reliable HRs of each source; then we
compared these parameters with the results of our numerical simulations.  In this way we
were able to constrain the possible range of values for each source spectral parameter, be
it temperature or photon--index, yielding a rough indication on the source type.

From the magnitudes of the selected counterparts (Bj for the GSC2.2 data and V for the WFI
ones), we obtained the corresponding optical fluxes and, then, the {\em
X--ray--to--optical} flux ratios.  In order to use this parameter for the source
characterization, we had to select a `reference $f_{X}/f_{opt}$ classification scheme'.
To this aim, we considered the results of the {\em Hamburg--RASS} optical identification
program \cite{Zickgraf}, which provides the typical ranges of values of the
$f_{X}/f_{opt}$ ratio for each class of celestial sources.  They were obtained using a
different X--ray energy range (0.1--2.4 keV instead of 0.3--10 keV), different spectral
models (specific for each class of sources instead of a fixed power--law) and, in the case
of the WFI data, a different optical filter (Johnson V instead of SERC Bj).  Therefore we
had to adapt the {\em Hamburg-RASS} results to the parameters used in our work
\cite{Chieregato}. Moreover, for a few of the WFI counterparts we had already a
suggested classification, based on the photometric data \cite{Hatziminaoglou}.

Based on the above procedure, we could divide our X--ray sources in three sets of data:
\begin{itemize}\vspace{-0.3 cm}
\item {\em Classified Sources}, which have a single optical counterpart and whose X--ray and optical parameters agrees to provide a reliable classification\vspace{-0.3 cm}
\item {\em Sources with candidate counterparts}, which have either more than one optical counterpart or a single one but whose X--ray and optical parameters do not yield a unique classification\vspace{-0.3 cm}
\item {\em Sources without candidate optical counterparts} within a 5$''$ radius\vspace{-0.3 cm}
\end{itemize}

\begin{table}[h]
\begin{center}
\caption{Results of the cross--correlation of the X--ray sources with the optical data}\label{statistics}
\begin{tabular}{|l|c|c|c|} \hline
EGRET field				&	\multicolumn{2}{|c|}{3EG J0616-3310}		&	3EG J1249--8330	\\ \hline
EPIC pointings				&	1		&	2-4		& 	5-8		\\ \hline
Total number of X--ray sources		& 	50		&	96		&	148		\\ \hline
Classified X--ray sources			& 	13 (26 \%)	&	24 (25 \%)	&	22 (15 \%)	\\ \hline
X--ray sources with candidate counterparts	&	10 (20 \%)	&	55 (57 \%)	&	32 (22 \%)	\\ \hline
Total number of candidate optical counterparts	&	26		&	126		&	61		\\ \hline
X--ray sources without candidate counterparts	&	27 (54 \%)	&	17 (18 \%)	&	94 (63 \%)	\\ \hline
\end{tabular}
\vspace{-1.5 cm}
\end{center}
\end{table}

Tab.~\ref{statistics} shows the number of sources for the three cases, together with the
total number of possible counterparts:  there we have considered the two {\em EGRET}
fields separately; moreover, we have also distinguished between
pointing which are covered by the GSC2.2 data only (1, 5-8) and pointings which are covered
by the WFI data too (2-4).  In the latter case there is a much lower percentage of
X--ray sources without candidate counterparts, in comparison with the former one; on
the other hand, both the percentage of X--ray sources with candidate counterparts and the
total number of candidate counterparts is very high.  These results are due to the deeper
optical coverage (see previous section).  This means that the missing counterparts are
fainter than V$\simeq$25.5 and Bj$\simeq$22.5 for, respectively, pointings 2-4 and
pointings 1, 5-8.

\begin{figure}[h]
\centering
\resizebox{\hsize}{!}{\includegraphics[width=2in,angle=-90]{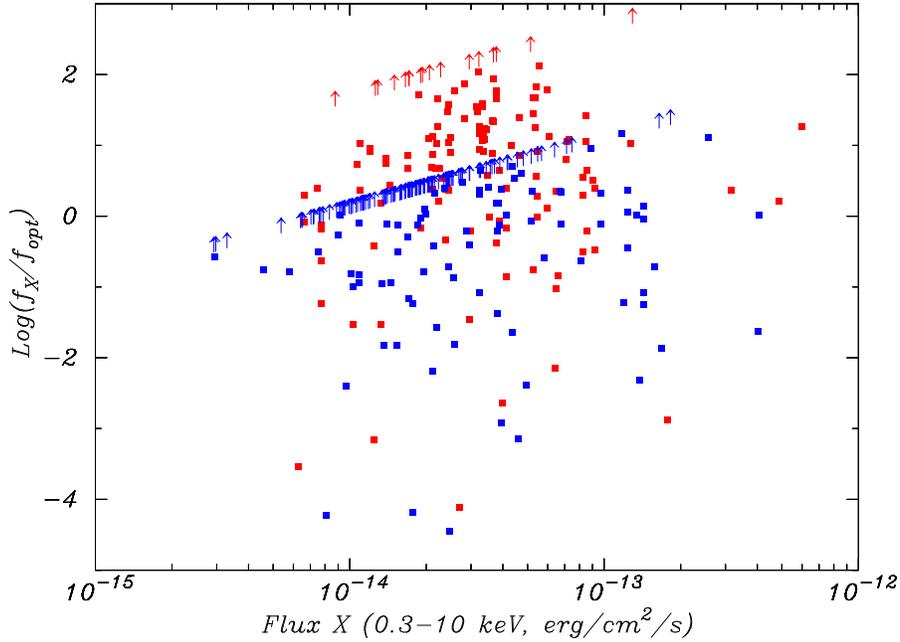}}
\caption{X--ray--to--optical flux ratio for the detected X--ray sources of our
sample.  Blue and red refer to {\em GSC2.2} and {\em WFI} counterparts,
respectively.  X--ray sources without optical counterparts (for which upper
limits have been used) are indicated with arrows.}\label{fluxratio}
\end{figure}

Fig.~\ref{fluxratio} shows the $f_{X}/f_{opt}$ values as a function of the X--ray fluxes
for all the X--ray sources of our sample, both with and without optical counterparts (for
sources without optical counterpart lower limits have been computed).  The plot shows that
there is a large spread in the $f_{X}/f_{opt}$ ratios at all values of the X--ray fluxes.

\vspace{-0.3 cm}
\begin{itemize}
\item When we consider only the sources with a candidate counterpart, we note that the
maximum $f_{X}/f_{opt}$ ratios are rather low:  $\simeq$15 for the sources with only a
GSC2.2 counterpart and $\simeq$130 for the sources with a WFI counterpart.  These values
are compatible with extragalactic identifications (AGN or cluster of galaxies) but not
with an INS one.\vspace{-0.3 cm}
\item When we include also the sources without optical counterpart we estimated the flux
ratios based on the limiting magnitudes (Bj$\simeq$22.5 or V$\simeq$25.5) and obtained
{\em lower limits} to the $f_{X}/f_{opt}$ values.  They range between 0.4 and 25 for the
GSC2.2 limit and between 45 and $\simeq$700 for the WFI limit:  this last value is very
interesting, since it approaches the expected X--ray--to--optical flux ratio for an INS
($\sim$1000).\vspace{-0.5 cm}
\end{itemize}
\vspace{-0.5 cm}

\section{Conclusions and perspectives}
\vspace{-0.3 cm}
Thanks to the {\em XMM-EPIC} observations, we have detected about 150 X--ray sources in
both the error--circle of 3EG J1249-8330 and 3EG J0616-3310.  Using survey optical data to
search for counterparts within 5$''$, we had a success rate of $\sim$ 40 \%, leaving
$\sim$ 60 \% of the X--ray sources unidentified down to a limiting magnitude
Bj$\simeq$22.5.  Ad hoc optical observations changed completely the scenario, since we
observed a candidate counterpart brighter than V$\simeq$25.5 for other 80 \% of the X--ray
sources.  The estimated lower limits of the X--ray--to--optical flux ratios of the
remaining sources are rather high and some of them potentially in agreement with the
expected value for an INSs.  Therefore, this sample of sources is very promising in order
to find the counterpart of the two unidentified EGRET sources and will be the object of
further, deeper investigations.

\end{article}

\vspace{-0.5 cm}
\begin{thebibliography}{}\vspace{-0.5 cm}
\bibitem[\protect\citeauthoryear{Baldi et al.}{2002}]{Baldi}
Baldi A. et al., 2002, Ap.J., 564, 190

\bibitem[\protect\citeauthoryear{Bignami and Caraveo}{1996}]{Bignami}
Bignami G.F. and Caraveo P.A., 1996, Ann. Rev. Astr. Astroph., 34, 331

\bibitem[\protect\citeauthoryear{Caraveo}{2001}]{Caraveo2001}
Caraveo P.A., 2001, Gamma--Ray Astrophysics, 4--6 April, 2001 Baltimore (MD), AIP Proceedings, 587, 641, astro-ph/0107370

\bibitem[\protect\citeauthoryear{Chieregato}{2004}]{Chieregato}
Chieregato M. et al., 2004, A\&A submitted

\bibitem[\protect\citeauthoryear{Hatziminaoglou et al.}{2000}]{Hatziminaoglou}
Hatziminaoglou E. et al., 2000, A\&A, 359, 9

\bibitem[\protect\citeauthoryear{La Palombara et al.}{2003}]{La Palombara}
La Palombara et al., 2003, Recent Res. Devel. Astronomy \& Astrophys., 1, 739

\bibitem[\protect\citeauthoryear{Mirabal and Halpern}{2001}]{Mirabal}
Mirabal N. and Halpern J.P., 2001, Ap.J., 547, L137

\bibitem[\protect\citeauthoryear{see Schirmer et al.}{2003 for the description of the data
processing}]{Schirmer}
Schirmer M. et al., 2003, A\&A, 407, 869

\bibitem[\protect\citeauthoryear{Str\"uder et al.}{2001}]{Struder}
Str\"uder L. et al., 2001, A\&A, 365, L18

\bibitem[\protect\citeauthoryear{Turner et al.}{2001}]{Turner}
Turner M.J.L. et al., 2001, A\&A, 365, L27

\bibitem[\protect\citeauthoryear{Zickgraf et al.}{2003}]{Zickgraf}
Zickgraf F.J. et al., 2003, A\&A, 406, 535

\end{thebibliography}
\end{document}